\documentstyle[twoside,fleqn,espcrc2]{article}

\input epsf
\def\spose#1{\hbox to 0pt{#1\hss}}
\def\lsim{\mathrel{\spose{\lower 3pt\hbox{$\mathchar"218$}}
 \raise 2.0pt\hbox{$\mathchar"13C$}}}
\def\gsim{\mathrel{\spose{\lower 3pt\hbox{$\mathchar"218$}}
 \raise 2.0pt\hbox{$\mathchar"13E$}}}

\newcommand{\AmS}{{\protect\the\textfont2
  A\kern-.1667em\lower.5ex\hbox{M}\kern-.125emS}}

\begin{document}

\begin{titlepage}

\begin{flushright}
CERN-TH/98-254\\
hep-ph/9808238
\end{flushright}

\vspace{1.5cm}

\boldmath
\begin{center}
\Large\bf Probing the CKM Angle $\gamma$ with $B\to\pi K$ Decays
\end{center}
\unboldmath

\vspace{1.2cm}

\begin{center}
Robert Fleischer\\
{\sl Theory Division, CERN, CH-1211 Geneva 23, Switzerland}
\end{center}

\vspace{1.3cm}

\begin{center}
{\bf Abstract}\\[0.3cm]
\parbox{11cm}{
The decays $B^\pm\to \pi^\pm K$ and $B_d\to\pi^\mp K^\pm$ were observed by 
the CLEO collaboration last year; they may play an important role to probe the 
angle $\gamma$ of the unitarity triangle of the CKM matrix. After a general 
parametrization of their decay amplitudes within the Standard Model, 
strategies to constrain and determine the CKM angle $\gamma$ with the help 
of the corresponding CP-conserving and CP-violating observables are briefly 
reviewed. The theoretical accuracy of these methods is limited by certain 
rescattering and electroweak penguin effects. It is emphasized that the 
rescattering processes can be included in the bounds on $\gamma$ by using 
additional experimental information on $B^\pm\to K^\pm K$ decays, and steps 
towards the control of electroweak penguins are pointed out. 
}

\end{center}

\vspace{1cm}

\begin{center}
{\sl Invited talk given at the\\
6th International Euroconference QCD 98\\
Montpellier, France, 2--8 July 1998\\
To appear in the Proceedings}
\end{center}

\vspace{1.5cm}

\vfil
\noindent
CERN-TH/98-254\\
August 1998

\end{titlepage}

\thispagestyle{empty}
\vbox{}
\newpage

\setcounter{page}{1}


\title{Probing the CKM angle $\gamma$ with $B\to\pi K$ decays}

\author{Robert Fleischer
\address{Theory Division, CERN, CH-1211 Geneva 23, Switzerland}}

\begin{abstract}
The decays $B^\pm\to \pi^\pm K$ and $B_d\to\pi^\mp K^\pm$ were observed by 
the CLEO collaboration last year; they may play an important role to probe the 
angle $\gamma$ of the unitarity triangle of the CKM matrix. After a general 
parametrization of their decay amplitudes within the Standard Model, 
strategies to constrain and determine the CKM angle $\gamma$ with the help 
of the corresponding CP-conserving and CP-violating observables are briefly 
reviewed. The theoretical accuracy of these methods is limited by certain 
rescattering and electroweak penguin effects. It is emphasized that the 
rescattering processes can be included in the bounds on $\gamma$ by using 
additional experimental information on $B^\pm\to K^\pm K$ decays, and steps 
towards the control of electroweak penguins are pointed out. 
\end{abstract}

\maketitle

\section{INTRODUCTION}\label{intro}

Among the central targets of the future dedicated $B$-physics experiments 
is the direct measurement of the three angles $\alpha$, $\beta$ and $\gamma$ 
of the usual, non-squashed, unitarity triangle of the 
Cabibbo--Kobayashi--Maskawa matrix (CKM matrix). From an experimental point 
of view, the determination of the angle $\gamma$ is particularly 
challenging, although there are several strategies on the market, 
allowing -- at least in principle -- a theoretically clean extraction of 
$\gamma$ (for a review, see for instance \cite{rev}). 

In order to probe this CKM angle in an experimentally feasible way, the 
decays $B^+\to\pi^+ K^0$, $B^0_d\to\pi^-K^+$ and their charge conjugates 
are very promising \cite{PAPIII}--\cite{wuegai}. Last year, the CLEO 
collaboration reported the observation of several exclusive $B$-meson 
decays into two light pseudoscalar mesons, including also these modes. 
So far, only results for the combined branching ratios
\begin{eqnarray}
\lefteqn{\mbox{BR}(B^\pm\to\pi^\pm K)\equiv}\nonumber\\
&&\frac{1}{2}\left[\mbox{BR}(B^+\to\pi^+K^0)+
\mbox{BR}(B^-\to\pi^-\overline{K^0})\right]\\
\lefteqn{\mbox{BR}(B_d\to\pi^\mp K^\pm)\equiv}\nonumber\\
&&\frac{1}{2}\left[\mbox{BR}(B^0_d\to\pi^-K^+)+
\mbox{BR}(\overline{B^0_d}\to\pi^+K^-)\right]
\end{eqnarray}
have been published at the $10^{-5}$ level with large experimental 
uncertainties \cite{cleo}. A particularly interesting situation arises, 
if the ratio
\begin{equation}\label{Def-R}
R\equiv\frac{\mbox{BR}(B_d\to\pi^\mp K^\pm)}{\mbox{BR}(B^\pm\to\pi^\pm K)}
\end{equation}
is found to be smaller than 1. In this case, the following allowed range
for $\gamma$ is implied \cite{fm2}:
\begin{equation}\label{gamma-bound1}
0^\circ\leq\gamma\leq\gamma_0\quad\lor\quad180^\circ-\gamma_0\leq\gamma
\leq180^\circ,
\end{equation}
where $\gamma_0$ is given by
\begin{equation}
\gamma_0=\arccos(\sqrt{1-R}).
\end{equation} 
Unfortunately, the present data do not yet provide a definite answer to the 
question of whether $R<1$. Since (\ref{gamma-bound1}) is complementary to 
the presently allowed range of 
$41^\circ\mathrel{\hbox{\rlap{\hbox{\lower4pt\hbox{$\sim$}}}\hbox{$<$}}}\gamma
\mathrel{\hbox{\rlap{\hbox{\lower4pt\hbox{$\sim$}}}\hbox{$<$}}}
134^\circ$ arising from the usual fits of the unitarity 
triangle \cite{burasHF97}, this bound would be of particular phenomenological
interest (for a detailed study, see \cite{gnps}). It relies on the following 
three assumptions:
\begin{itemize}
\item[i)] $SU(2)$ isospin symmetry can be used to derive relations between 
the $B^+\to\pi^+ K^0$ and $B^0_d\to\pi^-K^+$ QCD penguin amplitudes.
\item[ii)] There is no non-trivial CP-violating weak phase present in the
decay $B^+\to\pi^+ K^0$.
\item[iii)] Electroweak (EW) penguins play a negligible role in 
$B^+\to\pi^+ K^0$ and $B^0_d\to\pi^-K^+$.
\end{itemize}
Whereas (i) is on solid theoretical ground, provided the ``tree'' 
and ``penguin'' amplitudes of the $B\to\pi K$ decays are defined properly
\cite{bfm}, (ii) may be affected by rescattering processes of the kind
$B^+\to\{\pi^0K^+\}\to\pi^+K^0$, as was pointed out in 
\cite{FSI}--\cite{atso}. Concerning (iii), EW penguins may also play a 
more important role than is indicated by simple model calculations 
\cite{groro,neubert}. Consequently, in the presence of large rescattering
and EW penguin effects, strategies more sophisticated \cite{defan,rf-FSI}
than the ``na\"\i ve'' bounds sketched above, are needed to probe the CKM 
angle $\gamma$ with $B\to\pi K$ decays. Before turning to these methods,
let us first have a brief look at the corresponding decay amplitudes.

\boldmath
\section{THE $B\to\pi K$ DECAY AMPLITUDES}
\unboldmath

Within the Standard Model, the major contributions to $B^+\to\pi^+K^0$ and 
$B_d^0\to\pi^-K^+$ arise from QCD penguin topologies. In addition, 
annihilation topologies contribute to the former channel, while we have 
also colour-allowed $\bar b\to\bar uu\bar s$ tree-diagram-like topologies 
in the case of the latter decay. However, since these contributions are 
highly CKM-suppressed by $|V_{us}V_{ub}^\ast/(V_{ts}V_{tb}^\ast)|\approx0.02$ 
with respect to the QCD penguin contributions, the QCD penguins play the 
dominant role. 

Making use of the isospin symmetry of strong interactions, the QCD 
penguin topologies with internal top and charm quarks contributing to 
$B^+\to\pi^+K^0$ and $B_d^0\to\pi^-K^+$ can be related to each other, 
yielding the following amplitude relations (for a detailed discussion,
see \cite{bfm}):
\begin{eqnarray}
A(B^+\to\pi^+K^0)&\equiv&P\label{ampl-p}\\
A(B_d^0\to\pi^-K^+)&=&-\,\left[P+T+P_{\rm ew}\right],\label{ampl-n}
\end{eqnarray}
which play a central role to probe the CKM angle $\gamma$. Here the ``penguin''
amplitude $P$ is {\it defined} by the $B^+\to\pi^+K^0$ decay amplitude, 
$P_{\rm ew}\equiv-\,|P_{\rm ew}|e^{i\delta_{\rm ew}}$ is essentially due to 
electroweak penguins, and $T\equiv|T|e^{i\delta_T}e^{i\gamma}$ is usually 
referred to as a ``tree'' amplitude. However, due to a subtlety in the 
implementation of the isospin symmetry, the amplitude $T$ does not only 
receive contributions from colour-allowed tree-diagram-like topologies, but 
also from penguin and annihilation topologies \cite{bfm,defan}. The general 
expressions for the amplitudes $P$, $T$ and $P_{\rm ew}$, which are 
well-defined physical quantities, can be found in \cite{defan}. Let us here 
just note that we have
\begin{equation}
A(B^+\to\pi^+K^0)\propto\left[1+\rho\,e^{i\theta}e^{i\gamma}
\right]{\cal P}_{tc}\,,
\end{equation}
where
\begin{equation}\label{rho-def}
\rho\,e^{i\theta}=\frac{\lambda^2R_b}{1-\lambda^2/2}\left[1-\left(
\frac{{\cal P}_{uc}+{\cal A}}{{\cal P}_{tc}}\right)\right]
\end{equation}
is a measure of the strength of rescattering effects, as we will discuss in 
more detail in Section~\ref{Sec:resc}. In (\ref{rho-def}), $\theta$ is
a CP-conserving strong phase, $R_b\equiv|V_{ub}/(\lambda V_{cb})|=0.36\pm0.08$,
${\cal P}_{tc}\equiv|{\cal P}_{tc}|e^{i\delta_{tc}}$ and ${\cal P}_{uc}$ 
denote the differences of penguin topologies with internal top and charm and 
up and charm quarks, respectively, and ${\cal A}$ describes the annihilation 
topologies contributing to $B^+\to\pi^+K^0$.

In order to parametrize the $B^\pm\to \pi^\pm K$ and $B_d\to\pi^\mp K^\pm$
observables, it turns out to be very useful to introduce the quantities
\begin{equation}
r\equiv\frac{|T|}{\sqrt{\langle|P|^2\rangle}}\,,\quad\epsilon\equiv
\frac{|P_{\rm ew}|}{\sqrt{\langle|P|^2\rangle}}\,,
\end{equation}
with $\langle|P|^2\rangle\equiv(|P|^2+|\overline{P}|^2)/2$, as well
as the CP-conserving strong phase differences
\begin{equation}
\delta\equiv\delta_T-\delta_{tc}\,,\quad\Delta\equiv\delta_{\rm ew}-
\delta_{tc}\,.
\end{equation}
In addition to $R$ (see (\ref{Def-R})), also the ``pseudo-asymmetry'' $A_0$, 
which is defined by 
\begin{equation}
\frac{\mbox{BR}(B^0_d\to\pi^-K^+)-\mbox{BR}(\overline{B^0_d}\to
\pi^+K^-)}{\mbox{BR}(B^+\to\pi^+K^0)+\mbox{BR}(B^-\to\pi^-\overline{K^0})}\,,
\end{equation}
plays an important role to probe the CKM angle $\gamma$. Explicit expressions
for $R$ and $A_0$ in terms of the parameters specified above are given
in \cite{defan}.

\boldmath
\section{STRATEGIES TO CONSTRAIN AND DETERMINE THE CKM ANGLE $\gamma$}
\unboldmath

The observables $R$ and $A_0$ provide valuable information about the CKM
angle $\gamma$. If in addition to $R$ also the asymmetry $A_0$ can be 
measured, it is possible to eliminate the strong phase $\delta$ in the 
expression for $R$, and contours in the $\gamma$--$r$ plane can be fixed 
\cite{defan}, corresponding to a mathematical implementation of a simple 
triangle construction \cite{PAPIII}. In order to determine the CKM angle 
$\gamma$, the quantity $r$, i.e.\ the magnitude of the ``tree'' amplitude 
$T$, has to be fixed. At this step, a certain model dependence enters. 
In recent studies based on ``factorization'', the authors of 
Refs.~\cite{groro,wuegai} came to the conclusion that a future 
theoretical uncertainty of $r$ as small as ${\cal O}(10\%)$ may be 
achievable. In this case, the determination of $\gamma$ at future 
$B$-factories would be limited by statistics rather than by the uncertainty 
introduced through $r$, and $\Delta\gamma$ at the level of $10^\circ$ could 
in principle be achieved. However, since the properly defined amplitude $T$ 
does not only receive contributions from colour-allowed ``tree'' topologies, 
but also from penguin and annihilation processes \cite{bfm,defan}, it may be 
shifted sizeably from its ``factorized'' value so that 
$\Delta r={\cal O}(10\%)$ may be too optimistic. 

Interestingly, it is possible to derive bounds on $\gamma$ that do {\it not}
depend on $r$ at all \cite{fm2}. To this end, we eliminate again the
strong phase $\delta$ in the ratio $R$ of combined $B\to\pi K$ branching
ratios. If we treat now $r$ as a ``free'' variable, while keeping $\rho$
and $\epsilon$ fixed, we find that $R$ takes the following minimal 
value \cite{defan}: 
\begin{equation}\label{Rmin}
R_{\rm min}=\kappa\,\sin^2\gamma\,+\,
\frac{1}{\kappa}\left(\frac{A_0}{2\,\sin\gamma}\right)^2.
\end{equation}
In this expression, which is valid {\it exactly}, rescattering and EW 
penguin effects are described by
\begin{equation}
\kappa=\frac{1}{w^2}\left[\,1+2\,(\epsilon\,w)\cos\Delta+
(\epsilon\,w)^2\,\right]
\end{equation}
with
\begin{equation}
w=\sqrt{1+2\,\rho\,\cos\theta\cos\gamma+\rho^2}.
\end{equation}
An allowed range for $\gamma$ is related to $R_{\rm min}$, since values of
$\gamma$ implying $R_{\rm exp}<R_{\rm min}$, where $R_{\rm exp}$ denotes
the experimentally determined value of $R$, are excluded. The theoretical 
accuracy of these bounds on $\gamma$ is limited by rescattering and EW 
penguin effects, which will be discussed in the following two sections.
In the ``original'' bounds on $\gamma$ derived in \cite{fm2}, no information 
provided by $A_0$ has been used, i.e.\ both $r$ and $\delta$ were kept 
as ``free'' variables, and the special case $\rho=\epsilon=0$, i.e.\ 
$\kappa=1$, has been assumed, implying $\sin^2\gamma<R_{\rm exp}$. 
A measurement of $A_0\not=0$ allows us to exclude a range around $0^\circ$ 
and $180^\circ$, while the impact on the excluded range around $90^\circ$ 
is typically rather small. Interesting bounds on $\gamma$ can also
be obtained from $B_s\to K\overline{K}$ decays \cite{bsgam}.

\section{RESCATTERING PROCESSES}\label{Sec:resc}

In the formalism discussed above, rescattering processes are closely related
to the quantity $\rho$ (see (\ref{rho-def})), which is highly CKM-suppressed 
by $\lambda^2R_b\approx0.02$ and receives contributions from penguin 
topologies with internal top, charm and up quarks, as well as from 
annihilation topologies. Na\"\i vely, one would expect that annihilation 
processes play a very minor role, and that penguins with internal top quarks 
are the most important ones. However, also penguins with internal charm and
up quarks lead, in general, to important contributions \cite{LD-pens}. Simple
model calculations performed at the perturbative quark level do not indicate
a significant compensation of the large CKM suppression of $\rho$ through
these topologies. However, these crude estimates do not take into account 
certain rescattering processes \cite{FSI}--\cite{atso}, which may play an 
important role and can be divided into two classes \cite{bfm}:
\begin{itemize}
\item[i)] $B^+\to\{\overline{D^0}D_s^+,\,\overline{D^0}D_s^{\ast+},
\,\ldots\}\to\pi^+K^0$
\item[ii)]$B^+\to\{\pi^0K^+,\,\pi^0K^{\ast +},\,\ldots\}\to\pi^+K^0$,
\end{itemize}
where the dots include also intermediate multibody states. The rescattering
processes (i) can be considered as long-distance contributions to penguin
topologies with internal charm quarks and may affect BR$(B^\pm\to\pi^\pm K)$
significantly, while the final-state interaction (FSI) effects characterized 
by (ii) result in long-distance contributions to penguin topologies with 
internal up quarks and to annihilation topologies. They play a minor role 
for BR$(B^\pm\to\pi^\pm K)$, but may affect assumption (ii) listed in 
Section~\ref{intro}, thereby leading to a sizeable CP asymmetry, $A_+$, as 
large as ${\cal O}(10\%)$ in this mode \cite{gewe}--\cite{atso}. In a recent
attempt to evaluate rescattering processes of type (ii) with the help of
Regge phenomenology, it is found that $\rho$ may be as large as 
${\cal O}(10\%)$ \cite{fknp}. A similar feature is also present in other
approaches to deal with these FSI effects \cite{gewe,neubert}. Consequently, 
we have arguments that rescattering processes may in fact play an important 
role. 

A detailed study of their impact on the bounds on $\gamma$ arising from the 
$B^\pm\to \pi^\pm K$ and $B_d\to\pi^\mp K^\pm$ observables was performed 
in \cite{defan}. The FSI effects can be controlled through experimental data. 
A first step towards this goal is provided by the CP asymmetry $A_+$. In 
order to go beyond these constraints, $B^\pm\to K^\pm K$ decays -- the $SU(3)$ 
counterparts of $B^\pm\to \pi^\pm K$ -- play a key role, allowing us to 
include the rescattering processes completely in the contours in the 
$\gamma$--$r$ plane and the associated constraints on $\gamma$ 
\cite{defan,rf-FSI} (for alternative strategies, see \cite{bfm,fknp}). 

Since the ``short-distance'' expectation for the combined branching ratio
BR$(B^\pm\to K^\pm K)$ is ${\cal O}(10^{-6})$, experimental studies of 
$B^\pm\to K^\pm K$ appear to be difficult. These modes have not yet been 
observed, and only upper limits for BR$(B^\pm\to K^\pm K)$ are available 
\cite{cleo}. However, rescattering effects may enhance this 
quantity significantly, and could thereby make $B^\pm\to K^\pm K$ measureable 
at future $B$-factories \cite{defan,rf-FSI}. Another important indicator of 
large FSI effects is provided by $B_d\to K^+K^-$ decays \cite{groro-FSI}, 
for which stronger experimental bounds already exist \cite{cleo}.

\section{EW PENGUIN EFFECTS}

The modification of $R_{\rm min}$ through EW penguin topologies is described 
by $\kappa=1+2\,\epsilon\,\cos\Delta+\epsilon^2$. These effects are minimal
and only of second order in $\epsilon$ for $\Delta\in\{90^\circ,270^\circ\}$, 
and maximal for $\Delta\in\{0^\circ,180^\circ\}$. A detailed analysis can
be found in \cite{defan}. In the case of $\Delta=0^\circ$, which is favoured 
by ``factorization'', the bounds on $\gamma$ get stronger, excluding a larger 
region around $\gamma=90^\circ$, while they are weakened for 
$\Delta=180^\circ$. In the case of $B^+\to\pi^+K^0$ and $B_d^0\to\pi^-K^+$, 
EW penguins are ``colour-suppressed'' and estimates based on ``factorization'' 
typically give $\epsilon={\cal O}(1\%)$ (for a recent study, see \cite{AKL}). 
These crude estimates may, however, underestimate the role of these
topologies \cite{groro,neubert}. An improved theoretical description is 
possible by using the general expressions for the EW penguin operators and 
performing appropriate Fierz transformations, and a first step towards the 
experimental control of the relevant ``colour-suppressed'' EW penguin 
contributions is provided by the decay $B^+\to\pi^+\pi^0$ \cite{defan}. 
More refined strategies will certainly be developed in the future, when 
experimental data become available.

\section{CONCLUSIONS}

The decays $B^\pm\to \pi^\pm K$ and $B_d\to\pi^\mp K^\pm$ offer interesting 
strategies to probe the CKM angle $\gamma$. An accurate measurement of these 
modes, as well as of $B\to K\overline{K}$ and $B\to\pi\pi$ decays to control 
rescattering and EW penguin effects, is therefore an important goal of the 
future $B$-factories.

\end{document}